\documentclass[a4paper,11pt]{article}
\pdfoutput=1 
\usepackage{jheppub} 

\usepackage[T1]{fontenc} 

\pdfoutput=1
\usepackage{amsmath,amssymb}
\usepackage{amsfonts}
\usepackage{bm,bbm}
\usepackage{graphicx}
\usepackage{mathrsfs}
\usepackage{slashed}
\usepackage{booktabs}
\usepackage{tabu}
\usepackage[dvipsnames]{xcolor}
\usepackage[normalem]{ulem}
\usepackage{tikz}
\usepackage{soul}

\usepackage{hyperref}
\hypersetup{colorlinks,citecolor= blue,linkcolor= blue, urlcolor=blue}

\usepackage{xcolor}
\usepackage{ulem}
\usepackage{array}
\usepackage{verbatim}
\usepackage{epsfig}
\usepackage{multirow}

\newcommand{\be}{\begin{equation}}
\newcommand{\ee}{\end{equation}}
\newcommand{\ba}{\begin{array}}
\newcommand{\ea}{\end{array}}
\newcommand{\bea}{\begin{eqnarray}}
\newcommand{\eea}{\end{eqnarray}}

\usepackage{ulem,fancyvrb}
\usepackage{xcolor}

\title{Low-Energy Supernova Constraints on Lepton Flavor Violating Axions}

\author[a]{Zi-Miao Huang,}
\author[a]{Zuowei Liu}

\affiliation[a]{Department of Physics, Nanjing University, Nanjing 210093, China}

\emailAdd{zimiaohuang@smail.nju.edu.cn}
\emailAdd{zuoweiliu@nju.edu.cn}

\abstract{The extreme conditions within the supernova core, a high-temperature and high-density environment, create an ideal laboratory for the search for new physics beyond the Standard Model. Of particular interest are low-energy supernovae, characterized by their low explosion energies, which place strong constraints on the new-physics energy transfer from the core to the mantle. We compute low-energy supernova constraints on lepton-flavor-violating axions and axion-like particles that couple to both electrons and muons. For axion mass above the muon mass, the electron-muon coalescence and the axion decay are dominant production and reabsorption processes, respectively. We find that the low-energy supernovae provide the most stringent constraints on the axions in the mass range of $\sim (110,550)$ MeV, probing the coupling constant down to $g_{ae\mu} \simeq {\cal O}(10^{-11})$.}

\begin{document}

\maketitle
\flushbottom

\section{Introduction} 
\label{sec:introduction}

Axions are hypothetical particles proposed 
to resolve the strong CP problem in quantum chromodynamics 
\cite{Peccei:1977hh,Peccei:1977ur,Wilczek:1977pj,Weinberg:1977ma}. 
Beyond the original axion, 
a broader class of pseudo-scalar particles, 
commonly referred to as axion-like particles (ALPs), 
emerges naturally in many well-motivated extensions of the Standard Model (SM) 
\cite{Gelmini:1980re,Davidson:1981zd,Wilczek:1982rv,Svrcek:2006yi}. 
A particularly intriguing subclass of ALPs 
comprises those that exhibit lepton-flavor-violating (LFV) 
couplings to SM particles 
\cite{Wilczek:1982rv,Davidson:1981zd,Anselm:1985bp,Feng:1997tn,Bauer:2016rxs,Ema:2016ops,Calibbi:2016hwq,Choi:2017gpf,Chala:2020wvs,Bauer:2020jbp}. 
We will use ``axion'' and ``ALP'' interchangeably in this paper. 
The most stringent bounds on 
LFV ALPs with masses above the GeV scale  
arise from accelerator-based searches 
\cite{Endo:2020mev,Iguro:2020rby,
Davoudiasl:2021haa,Cheung:2021mol,Davoudiasl:2021mjy,Araki:2022xqp,Calibbi:2022izs,Batell:2024cdl,Calibbi:2024rcm}.  
In the sub-GeV mass regime, 
the leading bounds come from rare decay experiments 
\cite{Derenzo:1969za,Jodidio:1986mz,
Bryman:1986wn,Bilger:1998rp,TWIST:2014ymv,Bauer:2019gfk,
Cornella:2019uxs,PIENU:2020loi,Escribano:2020wua,Calibbi:2020jvd,Bauer:2021mvw,Knapen:2024fvh,Jho:2022snj,Knapen:2023zgi},
as well as astrophysical observations 
\cite{Calibbi:2020jvd,Zhang:2023vva,Li:2025beu}.

Recently, the supernova (SN) cooling bounds  
on LFV ALPs 
that couple to electrons and muons
have been investigated in 
Refs.~\cite{Calibbi:2020jvd,Zhang:2023vva,Li:2025beu}.
In this work, we extend the analysis by considering 
constraints from low-energy supernovae (LESNe), 
a subclass of underluminous Type II-P SNe, 
which are typically 10 to 100 times dimmer than 
ordinary core-collapse SNe (CCSNe). 
LESNe have been identified 
in both astronomical observations \cite{Chugai:1999en,Pastorello:2003tc,Pastorello:2009pt,10.1093/mnras/stu156,Pejcha:2015pca,Yang:2015ooa,Pumo:2016nsy,Murphy:2019eyu,Burrows:2020qrp,Yang:2021fka,Teja:2024cht} 
and SN simulations \cite{Kitaura:2005bt,Fischer:2009af,Melson:2015tia,Radice:2017ykv,Lisakov:2017uue,Muller:2018utr,Burrows:2019rtd,Stockinger:2020hse,Zha:2021bev}. 
LESNe can have an explosion energy as low as 
0.1 $\mathrm{B}=10^{50}$ erg, 
which has been found both in astronomical observations and 
in simulations \cite{Burrows:2020qrp}. 
For example, the reconstructed explosion energy of SN 1054 
is found to be around 0.1 B or less 
\cite{Yang:2015ooa,Stockinger:2020hse,Caputo:2022mah}.
In contrast to the conventional cooling bounds, 
which require the energy loss from new physics 
to be less efficient than that from neutrinos  
\cite{Raffelt:1996wa},  
the LESN constraints arise from the 
requirement that energy deposited in the SN mantle 
by new particles must remain below 
the LESN explosion energy, 0.1 B 
\cite{Falk:1978kf,Sung:2019xie,Caputo:2022mah}. 
Recent studies have demonstrated that 
LESNe can place powerful limits 
on a broad class of new physics scenarios 
\cite{Caputo:2022mah, Lella:2024gqc, Alda:2024cxn,
Chauhan:2023sci, Chauhan:2024nfa, Carenza:2023old,
Fiorillo:2024upk,Li:2024pcp,Fiorillo:2025yzf}.

For LFV ALPs that couple to electrons and muons and 
have masses below $\sim 100$ MeV, 
the leading constraints come from 
rare muon decay experiments 
\cite{Derenzo:1969za,Bilger:1998rp,Jodidio:1986mz,TWIST:2014ymv,PIENU:2020loi}. 
For LFV ALPs with masses above $\sim 100$ MeV, 
one of the leading constraints comes from the 
SN cooling bounds 
\cite{Li:2025beu}. 
The axion production channels inside the SN core include 
muon decay \cite{Calibbi:2020jvd}, 
axion bremsstrahlung 
\cite{Zhang:2023vva}, 
and 
electron-muon coalescence 
\cite{Li:2025beu}.  
For LFV ALPs with masses above $\sim 100$ MeV, 
the dominant axion production channel in the SN core 
is the electron-muon coalescence process \cite{Li:2025beu}. 
In this work, we focus on ALPs with masses above 
$\sim 100$ MeV and study the LESN constraints. 
By requiring that the energy deposition in the SN mantle through ALP decay to be less than the LESN explosion energy 0.1 B, we derive  constraints for  ALP masses in the range $m_a\sim(110,550)$ MeV. We find that the LESN constraints  probe new parameter space that is 
unconstrained by the SN cooling limits derived in Ref.~\cite{Li:2025beu}.

The rest of the paper is organized as follows. 
In section \ref{sec:model}, 
we introduce the LFV ALP  model. 
In section \ref{sec:energy:deposition}, 
we analyze the energy deposition in the SN mantle 
induced by ALPs produced in the SN core. 
In section \ref{sec:production:absorption}, 
we evaluate the ALP production and 
absorption rates. 
We present and discuss our results 
in section \ref{sec:results}, 
and summarize our main findings 
in section \ref{sec:conclusion}.

\section{ALP model} 
\label{sec:model}

We consider LFV ALPs that 
couple to electrons and muons  
via the following interaction Lagrangian: 
\begin{equation}
    \mathcal{L}_{\rm int} = \frac{g_{ae\mu}}{m_e^0+m_\mu}
    \bar e \gamma^\lambda \gamma_5 \mu 
    \partial_\lambda a + {\rm h.c.},
    \label{eq:lagrangian}
\end{equation}
where 
$a$ is the ALP field, 
$e$ is the electron field, 
$\mu$ is the muon field, 
$m_e^0=0.511$ MeV is the electron mass in vacuum, and 
$m_\mu=105.6$ MeV is the muon mass.  
The interaction Lagrangian given in Eq.~\eqref{eq:lagrangian} 
is equivalent to the following interaction
\begin{equation}
    \mathcal{L}_{\rm int} = -ig_{ae\mu} a 
    \bar e \gamma_5 \mu + {\rm h.c.},
    \label{eq:eff-lagrangian}
\end{equation}
if the fermions are on-shell 
\cite{Raffelt:1987yt,Lucente:2021hbp,Ferreira:2022xlw,Li:2025beu}. 
In this work, we use Eq.~\eqref{eq:eff-lagrangian} 
to compute the ALP 
production and reabsorption rates in the SN analysis.

\section{Energy deposition} 
\label{sec:energy:deposition}

LFV ALPs can be copiously produced in the SN core 
and subsequently reabsorbed in the SN mantle, 
leading to energy deposition in the mantle, 
which contributes to the SN explosion energy. 
Thus, to compute the LESN constraints, we first 
compute the energy deposition in the SN mantle 
caused by the reabsorption of ALPs
\cite{Caputo:2022mah}:
\begin{equation}
E_m
 = \Delta t \int_0^{R_{\mathrm{NS}}} d r \int_{m_a^{\prime}(r)}^{\infty} d E_a \frac{d L_a\left(r, E_a, t\right)}{d r \, d E_a} 
\left[
\exp  \left( \frac{r-R_{\mathrm{NS}}}{\lambda_a (r)}  \right)
-\exp \left( \frac{r-R_*}{\lambda_a (r)}  \right)
\right], 
\label{eq:energy-deposition}
\end{equation}
where 
$\Delta t = 3$ seconds is the typical SN explosion duration, 
$R_{\mathrm{NS}}$ is the core radius, 
$R_*$ is the radius of the progenitor star,   
$m_a^\prime = m_a / \mathrm{lapse}(r)$, which 
encodes gravitational redshift effects 
on the ALP mass $m_a$ through the lapse factor $\mathrm{lapse}(r)$ \cite{Caputo:2022mah}, 
$L_a\left(r, E_a, t\right)$ is the ALP luminosity 
with energy $E_a$ at radial coordinate $r$ 
and time $t$, 
and $\lambda_a(r)= {v_a}/{\Gamma_a}$ 
is the ALP mean free path 
with $v_a=\sqrt{E_a^2-m_a^2}/E_a$ 
and $\Gamma_a$ being the 
ALP velocity and decay width in the SN frame, respectively.

In our analysis, we use 
$R_{\mathrm{NS}} = 20$ km and 
$R_* = 3 \times 10^7$ km. 
Because typically $R_*$ is in the range of 
$\simeq (3-50) \times 10^7$ km \cite{Caputo:2022mah}, 
our choice 
represents a conservative estimate of the constraint. 
The ALP luminosity is computed via 
\begin{equation}
    \frac{d L_a\left(r, E_a, t\right)}{d r d E_a }= 4 \pi r^2 \, \mathrm{lapse}(r)^2 (1+2 v_r) E_a \frac{d^2 n_a}{dt \, dE_a}, \label{eq:luminosity}
\end{equation}
where $d^2 n_a/dtdE_a$ is the 
ALP production rate per unit volume and per unit energy, and 
$v_r$ is the velocity of the emitting medium in the radial direction. 
Because $v_r \ll 1$ \cite{Caputo:2022mah}, 
we neglect it in our analysis.

\section{ALP production and absorption} 
\label{sec:production:absorption}

In this work, we focus on LFV ALPs
with masses $m_a > m_\mu + m_e$. 
In general, 
LFV ALPs can be produced in the SN core via the 
following channels: 
muon decay, 
axion bremsstrahlung, and 
electron-muon coalescence 
\cite{Li:2025beu}. 
For $m_a > m_\mu + m_e$, 
the dominant production channel 
is the electron–muon coalescence process \cite{Li:2025beu}. 
Thus, in our analysis, 
we consider only ALP production through the electron-muon coalescence process, 
and neglect the muon decay and axion bremsstrahlung processes. 
For ALP reabsorption, 
we consider only the leading 
effects from the ALP decay process in this mass regime.

\begin{figure}[htbp]\centering
\includegraphics[width=0.3\textwidth]{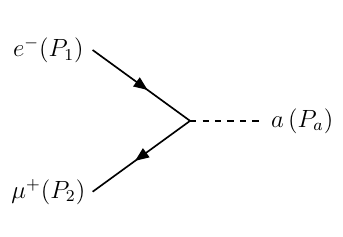}
\hspace{1cm}
\includegraphics[width=0.3\textwidth]{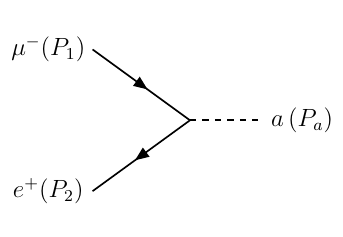}
\caption{The electron-muon coalescence processes 
for the ALP production: 
$e^- + \mu^+ \rightarrow a$ (left) and 
$\mu^- + e^+ \rightarrow a$ (right).
}
\label{fig:feynman}
\end{figure}

Fig.~\ref{fig:feynman} shows the two Feynman diagrams for 
the electron-muon coalescence process: 
$e^- + \mu^+ \rightarrow a$ and $\mu^- + e^+ \rightarrow a$.
The ALP production rate 
for $\mu^- + e^+ \rightarrow a$
is given by \cite{Li:2025beu} 
\begin{equation}
    \frac{d^2n_{a}}{dtdE_a} 
    (\mu^- + e^+ \rightarrow a)
    = \frac{\left|{\cal M}_{c}\right|^2}{32\pi^3}\int_{E_2^-}^{E_2^+} dE_2\ 
    f_\mu^- f_e^+, 
    \label{eq:ALP-production-rate:mu-e+}
\end{equation}
where $\left|{\cal M}_{c}\right|^2 = 2g_{ae\mu}^2[m_a^2 - (m_\mu-m_e)^2]$, 
$f_i^\mp$ is the Fermi-Dirac distribution for fermion (antifermion) $i$: 
\begin{equation}
f_i^\mp = \left[e^{(E_i \mp \mu_i)/T} + 1\right]^{-1}, \quad i = e,\mu.
\label{eq:antifermi-distribution}
\end{equation}
The upper and lower limits for the integral in Eq.~\eqref{eq:ALP-production-rate:mu-e+} 
are 
\begin{equation}
        E_2^\pm =\frac{E_a(m_a^2-m_1^2+m_2^2)}{2m_a^2} \pm \frac{\sqrt{E_a^2-m_a^2}}{2m_a^2}I,
\end{equation}
where $m_1$ ($m_2$) is the fermion (antifermion) mass and 
\begin{equation}
       I=\sqrt{(m_1^2-m_2^2-m_a^2)^2-4m_2^2m_a^2}.
\end{equation}
For the ALP production rate 
in the $e^- + \mu^+ \rightarrow a$ process,  
one changes 
$f_\mu^- f_e^+$
to 
$f_e^- f_\mu^+$ 
in Eq.~\eqref{eq:ALP-production-rate:mu-e+}: 
\begin{equation}
    \frac{d^2n_{a}}{dtdE_a} 
    (e^- + \mu^+ \rightarrow a) 
    = \frac{\left|{\cal M}_{c}\right|^2}{32\pi^3}\int_{E_2^-}^{E_2^+} dE_2\ 
    f_e^- f_\mu^+.  
    \label{eq:ALP-production-rate:e-mu+}
\end{equation}
Note that for both 
$\mu^- + e^+ \rightarrow a$
and 
$e^- + \mu^+ \rightarrow a$ processes, 
$P_1$ and $P_2$ denote the four-momenta of 
the initial state particle and anti-particle, 
respectively.

\begin{figure}[htbp]\centering
\includegraphics[width=0.5\textwidth]{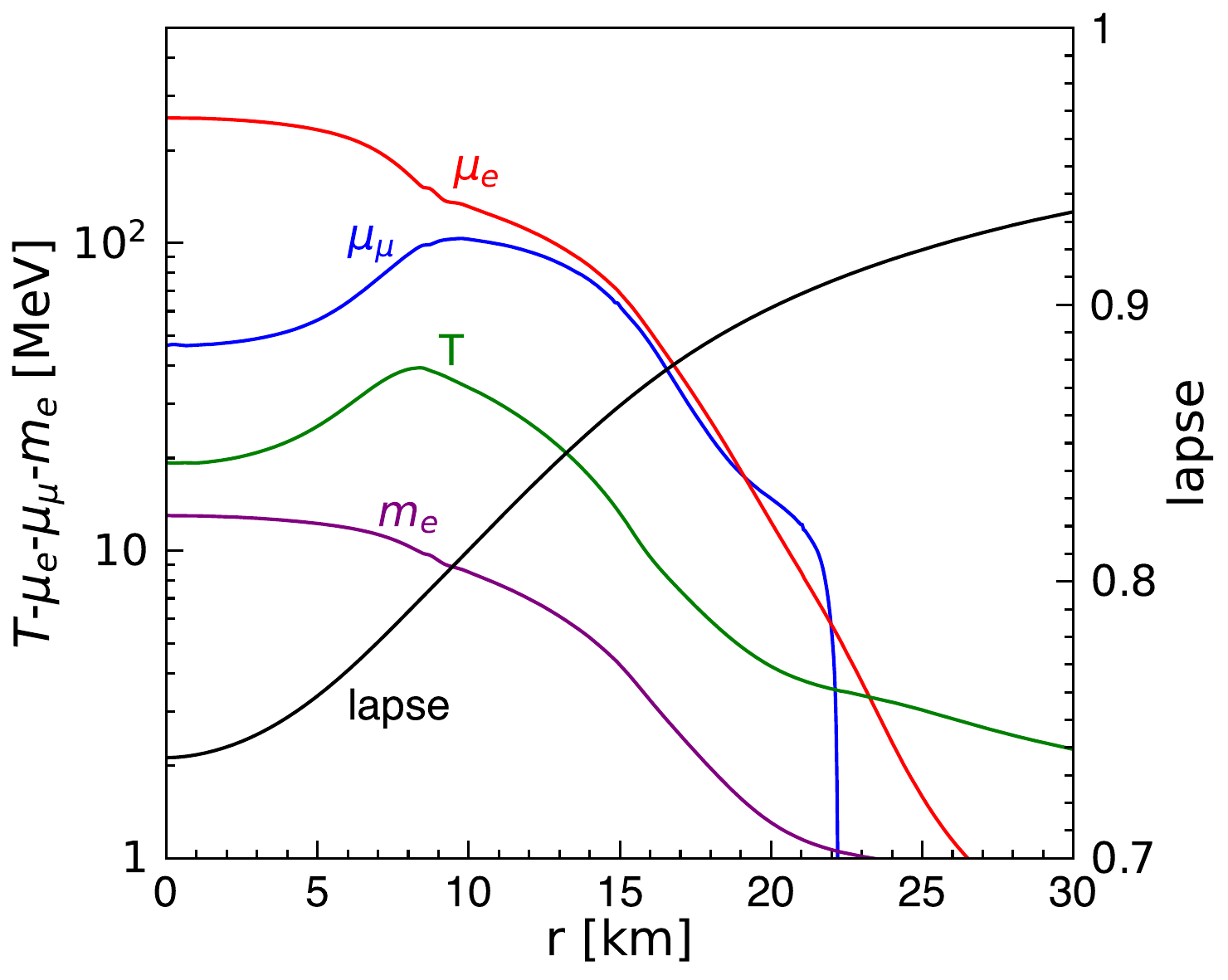}
\caption{The Garching profiles for the SFHo-18.8 SN model 
at 1 second postbounce \cite{garching-profile}: 
temperature $T$, 
the electron chemical potential $\mu_e$, 
the muon chemical potential $\mu_\mu$, 
the gravitational lapse factor, 
and the in-medium electron mass 
$m_e$ computed via Eq.~\eqref{eq:e-mass}.
}
\label{fig:profiles}
\end{figure}

To study LESN constraints, 
we follow Ref.~\cite{Caputo:2022mah} to 
adopt the Garching muonic model SFHo-18.8
\cite{Bollig:2020xdr,garching-profile}, 
the coldest model presented in Ref.~\cite{Bollig:2020xdr}. 
The SFHo-18.8 model has 
a peak temperature of 30-40 MeV and a final neutron star mass of 1.351 $M_\odot$ (baryonic) 
and 1.241 $M_\odot$ (gravitational) \cite{Caputo:2022mah}, 
which are compatible with predictions 
by theoretical models for LESNe 
\cite{Kitaura:2005bt,Janka:2007di,Hudepohl:2009tyy,
Fischer:2009af,Melson:2015tia,Radice:2017ykv,
Burrows:2019rtd,Glas:2018oyz,Muller:2018utr,Stockinger:2020hse,Zha:2021bev}.
Because the neutron star mass of the SFHo-18.8 model 
is close to the minimal neutron star mass 
expected in CCSNe \cite{Caputo:2022mah}, 
constraints derived with this model can be regarded as conservative estimates of LESN bounds. 
Fig.~\ref{fig:profiles} shows various profiles 
from the SFHo-18.8 model at 1 second postbounce, 
including temperature $T$, 
electron chemical potential $\mu_e$, muon chemical potential $\mu_\mu$,
and the gravitational lapse factor.
To accurately compute the ALP production rate   
via electron-muon coalescence 
and the ALP absorption rate via its inverse process,  
we use the in-medium electron mass, 
which receives significant corrections due to 
plasma effects in the SN core 
\cite{Braaten:1991hg,Lucente:2021hbp}: 
\begin{equation}
    m_e= \frac{m_e^0}{\sqrt{2}}+\sqrt{\frac{(m_e^0)^2}{2}+\frac{\alpha}{\pi}
    (\mu_e^2+\pi^2T^2 )}, 
    \label{eq:e-mass}
\end{equation}
where $\alpha=1/137$ is the fine-structure constant. 
We note that Eq.~\eqref{eq:e-mass} is an approximate expression,  
valid in the ultra-relativistic limit, 
which is appropriate for the SN analysis in this study.
We obtain 
the radial distribution of the in-medium electron mass $m_e$, 
by using the SN profiles of temperature $T$ and 
electron chemical potential $\mu_e$. 
The in-medium electron mass  
is found to decrease significantly 
from $\sim 12.9$ MeV at the core center ($r=0$)
to $\sim 1.3$ MeV at $r = 20$ km, 
as shown in Fig.~\ref{fig:profiles}.

\begin{figure}[htbp]\centering
\includegraphics[width=0.3\textwidth]{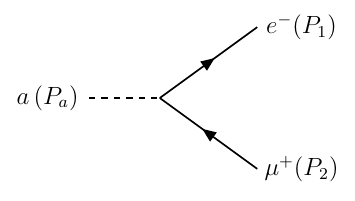}
\hspace{1cm}
\includegraphics[width=0.3\textwidth]{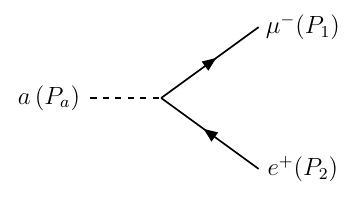}
\caption{The ALP decay processes: 
$a \rightarrow e^- + \mu^+ $ (left) and 
$a \rightarrow \mu^- + e^+$ (right).
}
\label{fig:feynman-ALP-decay}
\end{figure}

We next compute the ALP absorption rate.
For heavy ALPs with masses $m_a > m_\mu + m_e$, 
the dominant reabsorption processes are the 
ALP decay processes, 
$a \to e^\pm + \mu^\mp$, 
as shown in Fig.~\ref{fig:feynman-ALP-decay}, 
which are the inverse of the processes shown in 
Fig.~\ref{fig:feynman}. 
The absorption rate 
for the process of $a\to f$ is related to
the production rate via \cite{Li:2025beu} 
\begin{equation}
\Gamma_{a\to f}(E_a) 
= e^{(E_a-\mu_{a\to f}^0)/T} 
\frac{2\pi^2}{|{\bf p}_a|E_a}\frac{d^2n_a}{dt dE_a},
\label{eq:prod-abs} 
\end{equation}
where  
$|{\bf p}_a|=\sqrt{E_a^2-m_a^2}$, 
$d^2n_a/dt dE_a$ is the production rate 
for the corresponding electron-muon coalescence process,
and 
$\mu_{a\to f}^0 = \sum_{p} \mu_p$ with 
$\mu_p$ denoting the chemical potential 
of the particle $p$ in the final state. 
Thus, one has 
$\mu_{a\to e^- + \mu^+}^0 = \mu_{e} - \mu_{\mu}$  
for the process of $a \to e^- + \mu^+$,
and 
$\mu_{a\to \mu^- + e^+}^0 = \mu_{\mu} - \mu_{e}$ 
for the process of $a \to  \mu^- + e^+$,
where $\mu_e$ ($\mu_\mu$) denotes the 
electron (muon) chemical potential. 
We note that in regions where the ALP is in equilibrium, 
$\mu_{a\to f}^0$ becomes the chemical potential of the ALP 
and thus vanishes. In such cases, 
the electron chemical potential should be equal to 
the muon chemical potential, and 
one can no longer use the Garching profiles 
where the electron chemical potential 
is significantly different from 
the muon chemical potential. 
Equilibrium between the ALP and SM particles 
can be achieved 
in regions of parameter space where the coupling 
strength is sufficiently large. 
A detailed analysis of such cases 
will be presented in a future study 
\cite{LFV-LESN}.

\begin{figure}[htbp]\centering
\includegraphics[width=0.45\textwidth]{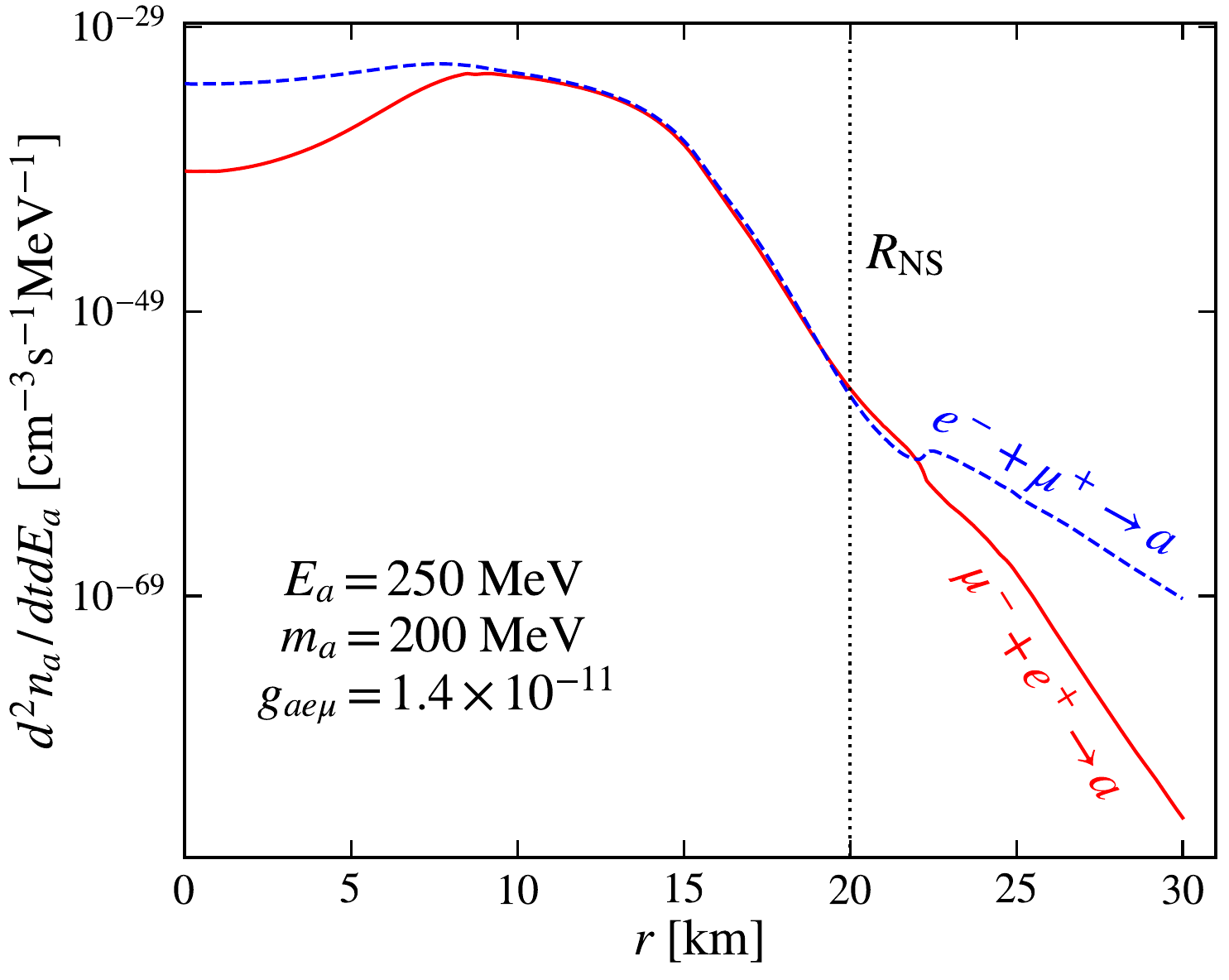}
\hspace{0.2cm}
\includegraphics[width=0.45\textwidth]{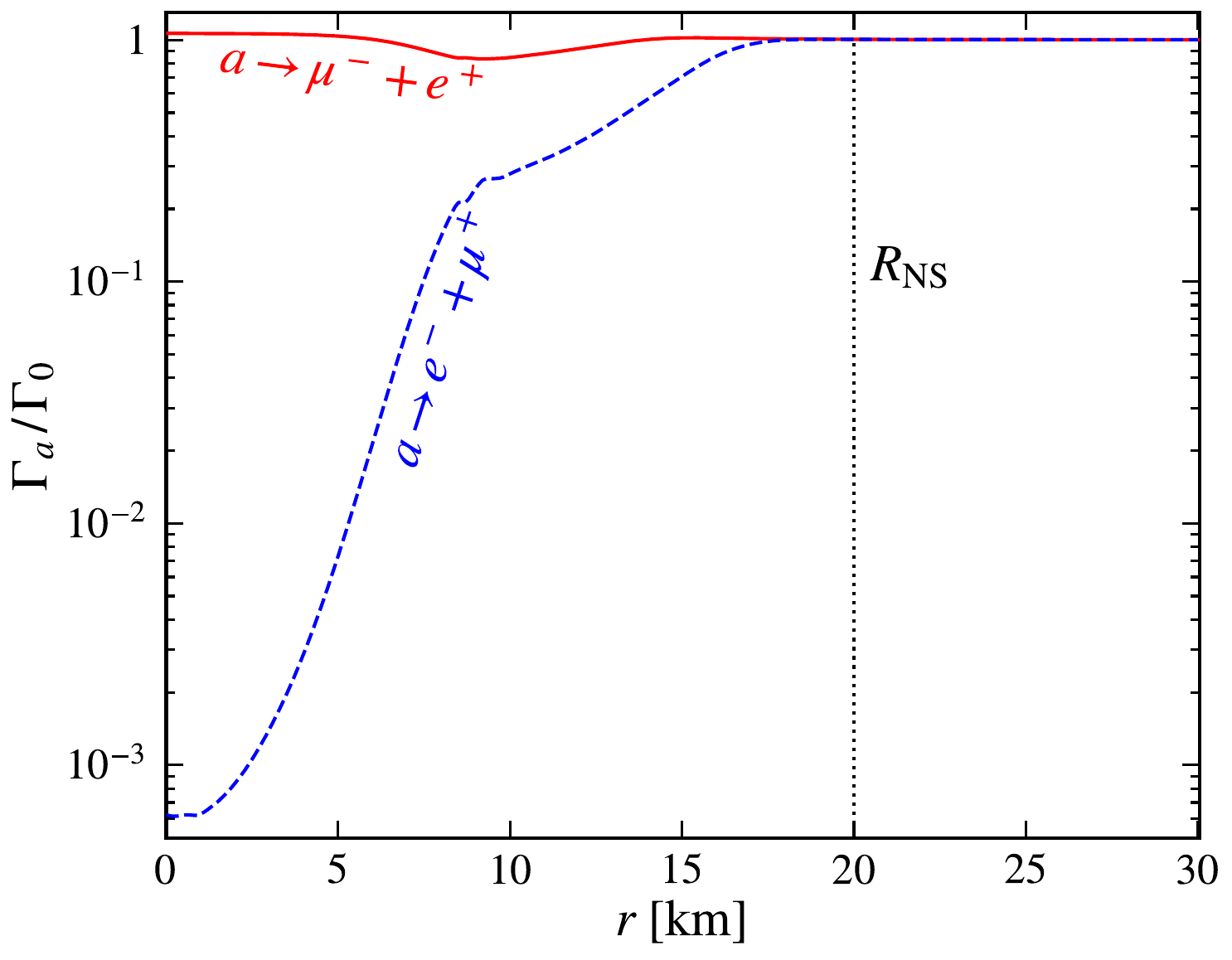}
\caption{
{\bf Left panel}:
ALP production rate as a function  of the radius $r$ 
in $e^- + \mu^+ \rightarrow a$ (blue dashed)  
and $\mu^- + e^+ \rightarrow a$ (red solid). 
{\bf Right panel}:
ALP absorption rate 
(normalized to its vacuum value, $\Gamma_0$) 
as function of radius $r$ 
for $a \rightarrow e^- + \mu^+$ (blue dashed)  
and $a \rightarrow \mu^- + e^+$ (red solid). 
For both  
panels, we use 
$E_a=250$ MeV, $m_a=200$ MeV, and $g_{ae\mu}=1.4 \times 10^{-11}$. 
$R_{\rm NS}=20$ km is 
the neutron star core radius (black dotted).} 
\label{fig:ALP-absorption-rate}
\end{figure}

\begin{figure}[htbp]\centering
\includegraphics[width=0.45\textwidth]{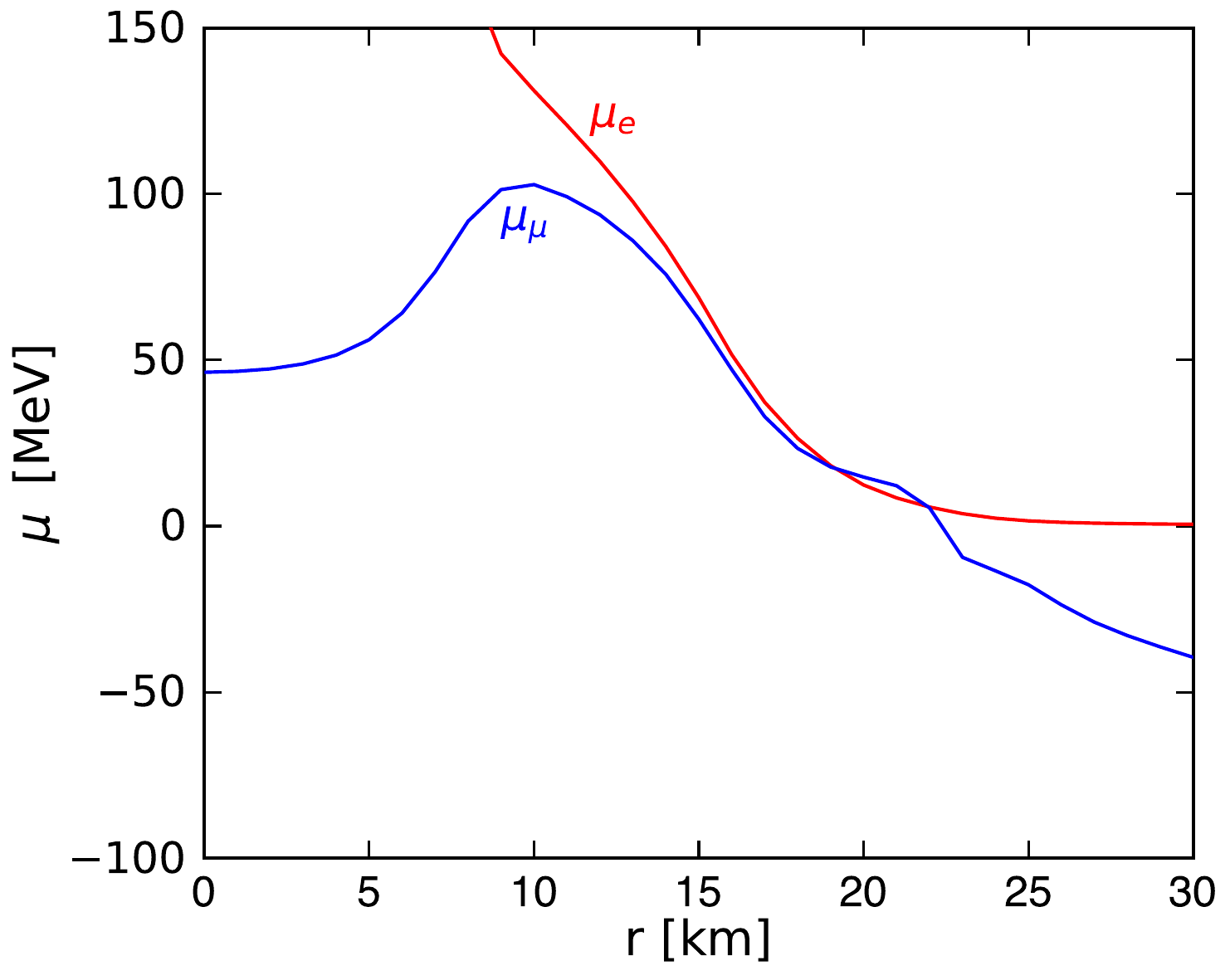}
\hspace{0.2cm}
\includegraphics[width=0.45\textwidth]{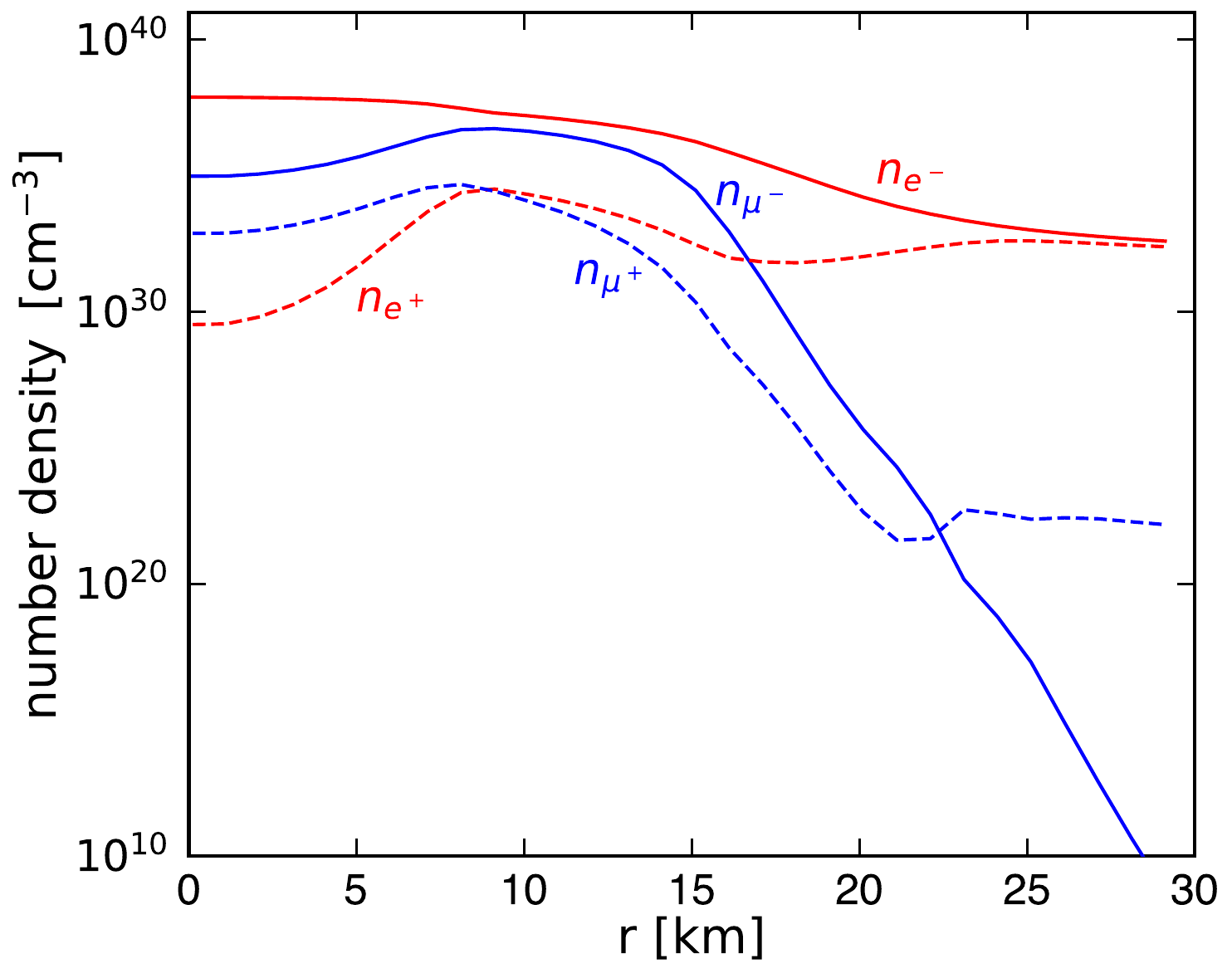}
\caption{
{\bf Left panel}:
Electron chemical potential $\mu_e$ (red) and muon chemical potential $\mu_\mu$ (blue) from the SFHo-18.8 model profiles \cite{garching-profile}.
{\bf Right panel}:
Number densities of $e^-$ (red-solid), $\mu^-$ (blue-solid), $e^+$ (red-dashed), and $\mu^+$ (blue-dashed) calculated using Eq.~\eqref{eq:number-density}.} 
\label{fig:number-density}
\end{figure}

Fig.~\ref{fig:ALP-absorption-rate} shows the ALP production rates 
for the two electron-muon coalescence processes 
shown in Fig.~\ref{fig:feynman}, 
and the absorption rates for the two ALP decay processes 
shown in Fig.~\ref{fig:feynman-ALP-decay}, 
as a function of the radius $r$, 
for the benchmark model point: 
$(E_a, m_a, g_{ae\mu})=(250 \, \mathrm{MeV}, 200 \, \mathrm{MeV}, 1.4 \times 10^{-11})$.
For $10  \lesssim  r \lesssim 22 \, \mathrm{km}$, 
the two electron-muon coalescence processes 
yield similar production rates; 
for $r \lesssim 10$ km and $r \gtrsim 22$ km, 
the $e^- + \mu^+ \to a$ process 
has a larger production rate than 
the $\mu^- + e^+ \to a$ process. 
To understand this behavior, we compute the number density 
\begin{equation}
    n_{\ell^\mp} (r) = \frac{1}{\pi^2} \int_0^{\infty} \mathrm{d}p \, p^2 \, f_\ell^\mp(r), 
    \quad \ell = e,\mu, 
    \label{eq:number-density}
\end{equation}
where $f_\ell^\mp$ is given in  
Eq.~\eqref{eq:antifermi-distribution}. 
As shown in the right panel of Fig.~\ref{fig:number-density}, 
in the $r \lesssim 10\,\mathrm{km}$ region we have 
$n_{e^-} > n_{\mu^-}$ and $n_{\mu^+} > n_{e^+}$, 
while in the $r \gtrsim 22\,\mathrm{km}$ region we have 
$n_{e^-} > n_{e^+}$ and $n_{\mu^+} > n_{\mu^-}$. 
Therefore, in both regions, 
$n_{e^-} n_{\mu^+} > n_{\mu^-} n_{e^+}$, 
which explains why  
the $e^- + \mu^+ \to a$ channel dominates over the $\mu^- + e^+ \to a$ 
channel there. 
Note that $n_{\mu^+} > n_{\mu^-}$ in the $r \gtrsim 22\,\mathrm{km}$ region 
arises because the muon chemical potential becomes 
negative in that region, 
as shown in the left panel of Fig.~\ref{fig:number-density}.

The ALP absorption rates shown in 
Fig.~\ref{fig:ALP-absorption-rate} 
are normalized to the decay width in the vacuum, $\Gamma_0$, which is 
\begin{equation}
    \Gamma_0
    = \frac{g_{ae\mu}^2}{8\pi E_a m_a^2}
    \left[m_a^2-(m_\mu-m_e^0)^2\right] \sqrt{(m_\mu^2-\left( m_e^0 \right)^2-m_a^2)^2-4\left( m_e^0 \right)^2m_a^2}.
\end{equation}
For $r\lesssim 15$ km, 
the decay channel $a \rightarrow \mu^- + e^+$  
dominates over 
$a \rightarrow e^- + \mu^+$.  
This pronounced disparity arises primarily 
from the exponential factor in Eq.~\eqref{eq:prod-abs}.  
At these radii, the electron chemical potential $\mu_e$ 
is significantly larger 
than the muon chemical potential $\mu_\mu$, 
as shown in  Fig.~\ref{fig:profiles}. 
This generates 
a large and positive $\mu_{a\to f}^0$ 
for the $a \to e^- + \mu^+$ channel, 
and an equally large and negative 
$\mu_{a\to f}^0$ for the $a \to \mu^- + e^+$ channel. 
This sign difference 
leads to an exponential 
suppression of the $a \to e^- + \mu^+   $ decay rate  
relative to $a \to \mu^-  +  e^+$, 
despite the former having a larger production rate. 
Note that the above discussion 
assumes that ALPs are not in equilibrium. 
As we discussed previously following Eq.~\eqref{eq:prod-abs}, 
in regions of parameter space where the coupling strength is sufficiently large,  
equilibrium between the ALP and SM particles can be achieved, 
leading to equal electron and muon chemical potentials. 
In such cases, one can no longer use the current Garching profiles. 
A detailed analysis of such cases will be presented in a future study \cite{LFV-LESN}.

\section{Results} 
\label{sec:results}

We compute the energy deposition in the SN mantle $E_m$ via Eq.~\eqref{eq:energy-deposition} 
for heavy 
LFV ALPs with masses $m_a > m_\mu + m_e$. 
In this mass regime, we consider only the dominant production process 
through the $e$-$\mu$ coalescence and the primary reabsorption term for the ALP decay process. 
By requiring the energy deposition in the SN mantle to be less than the LESN explosion energy, 
$E_m \le 0.1$ B, 
we derive the constraints on the LFV ALPs.

\begin{figure}[htbp]\centering
\includegraphics[width=0.5\textwidth]{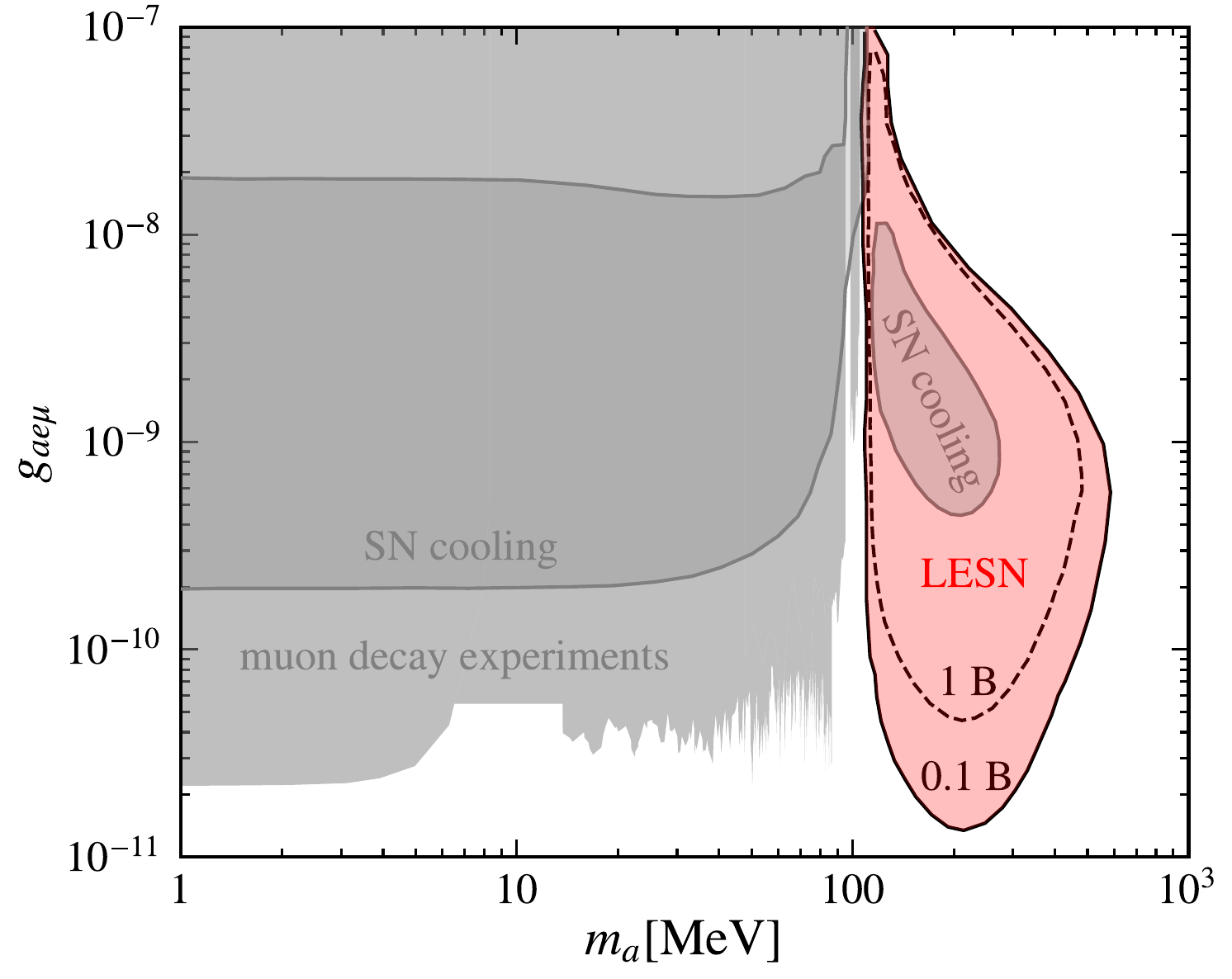}
\caption{The LESN constraints on LFV ALPs (red-shaded region) 
for $E_m \le 0.1$ B (solid). The case of $E_m \le 1$ B (dashed) is also shown.
Gray regions are 
the SN cooling constraints \cite{Li:2025beu}, 
and constraints from rare muon decay experiments 
\cite{Derenzo:1969za,Bilger:1998rp,Jodidio:1986mz,TWIST:2014ymv,PIENU:2020loi}, 
which are taken from Ref.~\cite{Li:2025beu}.} 
\label{fig:constraints}
\end{figure}

Fig.~\ref{fig:constraints} shows 
the LESN constraints on LFV ALPs 
with the requirement of $E_m \le 0.1$ B,  
along with other existing experimental constraints, 
including 
those that arise from the rare muon decay experiments 
\cite{Derenzo:1969za,Bilger:1998rp,Jodidio:1986mz,TWIST:2014ymv,PIENU:2020loi}, 
and 
SN cooling limits 
\cite{Li:2025beu}. 
For ALPs with 
masses $m_a \lesssim 105~\mathrm{MeV}$, 
the SN cooling limits are primarily due to two 
ALP production channels in the SN core: 
muon decay and axion bremsstrahlung. 
For ALPs with 
masses $m_a \gtrsim 105~\mathrm{MeV}$, 
the dominant ALP production channel in the SN core is 
the electron-muon coalescence \cite{Li:2025beu}. 
For $m_a \lesssim 105~\mathrm{MeV}$, 
the most stringent limits arise from the rare muon decay experiments, 
which are stronger than SN cooling limits. 
For $m_a \gtrsim 105~\mathrm{MeV}$, 
the previously most important constraints came 
from SN cooling limits, 
where rare muon decay experiments are kinematically forbidden. 
In Fig.~\ref{fig:constraints}, we also show the LESN constraints
for the energy deposition of $E_m \le 1$ B, 
which provide weaker  limits than the case of $E_m \le 0.1$ B.

We find that LESN constraints 
provide the most stringent constraints 
in the ALP mass range 
$m_a\sim(110,550)$ MeV, 
extending the parameter space excluded by 
the SN cooling constraints  \cite{Li:2025beu} 
to a much larger exclusion region.
In particular, the LESN constraints  probe 
the coupling $g_{ae\mu}$ 
down to $\sim  10^{-11}$ for $m_a \approx 200$ MeV, 
which is one order of magnitude stronger    
than the SN cooling constraints given by 
Ref.~\cite{Li:2025beu}. 
Note that the LESN constraints are 
manifested as 
an exclusion region, where the 
upper boundary arises 
from significant reabsorption effects 
at strong couplings, while 
the lower boundary results from 
suppressed ALP production at weak couplings. 
Also note that for $m_a\le 105 $ MeV, 
ALP production through electron-muon coalescence becomes kinematically forbidden 
so that the low-mass region is not accessible via  
electron-muon coalescence in this analysis.

\section{Conclusion} 
\label{sec:conclusion}

In this paper, we compute the LESN constraints on  LFV ALPs with masses $m_a > m_\mu + m_e$, by requiring the energy deposition in the SN mantle to be less than the LESN explosion energy $0.1$ B. For this mass range, we consider the dominant ALP production channel through the $e$-$\mu$ coalescence process  \cite{Li:2025beu} and the subsequent energy deposition from ALP decay. 
We find that the LESN constraints 
probe a much larger parameter space than the 
SN cooling limits, 
extending to the 
previously unexplored parameter space for ALP masses between 110 MeV and 550 MeV. 

\begin{acknowledgments}
We thank Changqian Li, Yonglin Li, Wenxi Lu, and Zicheng Ye 
for discussions. 
We thank Hans-Thomas Janka 
for providing the SN profiles 
used for numerical calculations. 
The work is supported in part by the 
National Natural Science Foundation of China under Grant 
No.\ 12275128. 
\end{acknowledgments}

\normalem
\bibliography{ref.bib}{}
\bibliographystyle{utphys28mod}
\end{document}